# Does Military Expenditure Impede Sustainable Development? Empirical Evidence from NATO Countries

Askeri Harcamalar Sürdürülebilir Kalkınmayı Engelliyor mu? NATO Ülkelerinden Ampirik Kanıtlar

Emre AKUSTA*

* Asst. Prof., Kırklareli University, Faculty of Economics And Administrative Sciences, Kırklareli, Türkiye
e-mail: emre.akusta@klu.edu.tr
ORCID: 0000-0002-6147-5443

**Abstract**

This study analyzes the impact of military expenditures on sustainable development in NATO countries. The analysis utilizes annual data for the period between 1995 and 2019. In this study, the Durbin-Hausman panel cointegration test is used to analyze the cointegration relationship between the variables and the Panel AMG estimator is used to estimate the long-run coefficients. The results of the AMG estimator show that military expenditures and industrial production index have a negative effect on sustainable development in NATO countries, while foreign direct investments have a positive effect. The impact of primary energy consumption is negative and less significant than the other negative impacts. The study also analyzes how the impact of military expenditures on sustainable development varies across countries. This analysis reveals the significant differences in the direction, significance, and coefficient size of the relationship among different countries. These findings suggest that the impact of military expenditures on sustainable development varies across countries. Therefore, countries should develop policies to ensure sustainable development by considering their specific dynamics.

**Keywords:** Military Expenditures, Sustainable Development, NATO, Panel Data Analysis, Panel AMG Estimator

**Özet**

Bu çalışma, NATO ülkelerinde askeri harcamaların sürdürülebilir kalkınma üzerindeki etkilerini analiz etmektedir. Analizde 1995-2019 dönemi için yıllık veriler kullanılmıştır. Çalışmada değişkenler arasındaki eş bütünleşme ilişkisini araştırmak için Durbin-Hausman panel cointegration test; uzun dönem katsayıları tahmin etmek amacıyla ise Panel AMG tahmincisi kullanılmıştır. AMG tahmincisi sonuçları gösteriyor ki, NATO ülkelerinde askeri harcamalar ve sanayi üretim endeksi, sürdürülebilir kalkınma üzerinde negatif bir etki sergilerken, yabancı doğrudan yatırımlar pozitif bir etki yaratmıştır. Enerji tüketiminin etkisi ise negatif olup, diğer negatif etkilere göre daha az belirgindir. Çalışmada ayrıca askeri harcamaların sürdürülebilir kalkınma üzerindeki etkisinin ülke özelinde nasıl değişkenlik gösterdiği incelendi. Analizler, farklı ülkelerdeki ilişkinin yönü, anlamlılığı ve katsayı büyüklüğü açısından önemli farklılıklar ortaya koymuştur. Bu bulgulara göre, askeri harcamaların sürdürülebilir kalkınma üzerindeki etkileri ülkeden ülkeye değişmektedir. Bu nedenle ülkeler, kendine özgü dinamiklerini göz önünde bulundurarak sürdürülebilir kalkınmalarını sağlamak için politikalar geliştirmelidir.

**Anahtar Kelimeler:** Askeri Harcamalar, Sürdürülebilir Kalkınma, NATO, Panel Veri Analizi, Panel AMG Tahmincisi

**Introduction**

One of the most prominent features of the 21st century is the ongoing power struggles in the global economic and political arena. While these power struggles cause countries to increase their military expenditures, they also pose serious obstacles to achieving sustainable development goals. Within this context, understanding and analyzing how military expenditures affect sustainable development is of great importance for both academia and policymakers.

Sustainable development is a development model that aims to balance between economic growth, social welfare, and environmental sustainability. The 1987 Brundtland Report defines it as a development process that "meets the needs of current generations without compromising the ability of future generations to meet their own needs".[1] The Sustainable Development Goals (SDGs), adopted by the United Nations (UN) in 2015, have been accepted worldwide with 17 main goals and 169 sub-goals as the embodiment of this concept. The SDGs consist of a series of interconnected plans adopted by the countries and regions of the world, aiming to achieve sustainable development by 2030. The SDGs are a continuation of the Millennium Development Goals (MDGs)[2] and aim to continue the agenda set by the MDGs.[3]

Military expenditure refers to the funds allocated to a country's defense budget. It ranges from military personnel salaries, arms and ammunition purchases, research and development activities, military operations, and maintenance costs. The Stockholm International Peace Research Institute (SIPRI) reports that military expenditures worldwide have increased steadily in recent years and account for a significant share of the global Gross Domestic Product (GDP). Given that resources are scarce, public expenditures such as military expenditures can affect the quality of budget allocations for SDGs and, thus, the likelihood of their implementation.[4]

The impact of military expenditures on sustainable development can be analyzed in two main categories: direct and indirect. Direct effects include the burden of military expenditures on the public budget and the reduction in expenditures in critical areas for sustainable development, such as health, education, and infrastructure. Indirect effects include the impact of military expenditure on economic growth, income distribution, and environmental sustainability. Direct effects are particularly significant in developing countries where high military expenditures prevent allocating public resources to areas such as education, health, and social services. For example, according to UNICEF data, many African countries have had to reduce investments in education and health while increasing military expenditures. This restricts access to education and health services, negatively impacting human capital development and sustainable development in the long run. On the other hand, the indirect effects of military expenditures manifest themselves in economic and environmental dimensions. In the economic dimension, high military expenditures lead to inefficient use of resources and negatively impact economic growth. Military expenditures may also increase aggregate demand by substituting non-defense public expenditures. However, this increase does not lead to sustainable growth in the long run. On the contrary, directing resources to the defense sector may slow economic growth by limiting innovation and productivity growth.[5] Regarding the environmental dimension, military activities and the weapons and ammunition used for these activities cause environmental pollution and the depletion of natural resources. For example, heavy vehicles and explosives used during military exercises can cause permanent damage to soil and water resources. In addition, environmental destruction in war and conflict zones leads to the degradation of ecosystems and the reduction of biodiversity.[6] There is a broad consensus in the literature that military activities cause significant environmental damage.[7] Military airplanes, helicopters, ships, tanks, and other military machinery and equipment consume large amounts of fossil and nuclear fuels, which is one factor contributing to environmental problems.[8] The UN's Intergovernmental Panel on Climate Change has emphasized that increasing carbon dioxide emissions due to high levels of fossil fuel use will cause climatic changes.

---

1 Robert H. Cassen, "Our Common Future: Report of the World Commission on Environment and Development", *International Affairs,* 64:1, 1987, p. 126.
2 Marta Lomazzi, Bettina Borisch and Ulrich Laaser, "The Millennium Development Goals: Experiences, Achievements and What's Next", *Global Health Action,* 7:1, 2014, Vol. 23695, p.p 109-146.
3 Pamela S. Chasek et al., "Getting to 2030: Negotiating the Post-2015 Sustainable Development Agenda", *Review of European, Comparative & International Environmental Law,* 25: 1, 2016, pp. 5-14.
4 Omar A. Guerrero and Gonzalo Castañeda, "How Does Government Expenditure Impact Sustainable Development? Studying the Multidimensional Link between Budgets and Development Gaps", *Sustainability Science,* 17: 3, 2022, pp. 987-1007; SIPRI Military Expenditure Database (Stockholm International Peace Research Institute, 2024).

Theoretical and empirical studies examining the impact of military expenditure on sustainable development generally emphasize the negative impacts of such expenditures. Peace Economics Theory argues that military expenditures are destructive and unproductive rather than peaceful and productive uses of resources. Empirical studies show that high military expenditures worsen socioeconomic indicators in the long run by reducing health, education, and infrastructure investments. However, results vary from region to region and depending on the scope of the study. Therefore, this study analyzes the impact of military expenditures on sustainable development in NATO countries. NATO countries represent a large share of military spending worldwide. These countries usually have developed economies and play a decisive role in global security policies. This makes NATO countries' military expenditures strategically important for us to understand global military spending trends. Moreover, the economic and political diversity among NATO countries also allows for a comparative assessment of the impacts of military expenditures on different economic and social structures.

This study contributes to the literature in at least four ways: (1) To the best of our knowledge, there is no empirical study investigating the impact of military spending on sustainable development for NATO countries. This study aims to fill this gap in the literature. (2) The impact of military expenditures on sustainable development is analyzed in a way that includes not only economic but also social and environmental dimensions. Therefore, this study analyzes the impact of military expenditures more holistically. (3) The modeling in this study is conducted for both the NATO alliance and the member countries, while the differences between countries are not ignored. In this way, it is possible to analyze how the effects of military expenditures differ across countries. (4) This study employs second-generation panel data techniques that consider cross-sectional dependence. These methods provide more robust and valid results.

---

5 Selahattin Bekmez and M. Akif Destek, "Savunma Harcamalarında Dışlama Etkisinin İncelenmesi: Panel Veri Analizi", *Siyaset, Ekonomi ve Yönetim Araştırmaları Dergisi,* 3: 2, 2015, p. 91-110.
6 Kenneth A. Gould, "The Ecological Costs of Militarization", *Peace Review,* 19: 3, 2007, p. 331–34; Gregory Hooks and Chad L. Smith, "The Treadmill of Destruction: National Sacrifice Areas and Native Americans", *American Sociological Review,* 69: 4, 2004, p. 558-575.
7 Aliya Zhakanova Isiksal, "Testing the Effect of Sustainable Energy and Military Expenses on Environmental Degradation: Evidence from the States with the Highest Military Expenses", *Environmental Science and Pollution Research,* 28: 16, 2021, Vol. 20487, pp. 75-98; Andrew K. Jorgenson, Brett Clark, and Jeffrey Kentor, "Militarization and the Environment: A Panel Study of Carbon Dioxide Emissions and the Ecological Footprints of Nations, 1970-2000", *Global Environmental Politics,* 10: 1, 2010, pp. 7-29; Sakiru Adebola Solarin, Usama Al-mulali, and Ilhan Ozturk, "Determinants of Pollution and the Role of the Military Sector: Evidence from a Maximum Likelihood Approach with Two Structural Breaks in the USA", *Environmental Science and Pollution Research,* 25:31, 2018, Vol. 30949, pp. 45-61.
8 Melike Bildirici, "CO2 Emissions and Militarization in G7 Countries: Panel Cointegration and Trivariate Causality Approaches", *Environment and Development Economics* 22: 6, 2017, pp. 71-91; David Naguib Pellow, *Resisting Global Toxics: Transnational Movements for Environmental Justice*, MIT Press, 2007.

The rest of the paper is organized as follows: Section 1 presents NATO countries' military expenditures and sustainable development outlook, Section 2 presents a literature review, Section 3 presents data and methodology, Section 4 presents results and discussion, and Section 5 presents conclusions.

## 1. NATO Countries' Military Expenditures and Sustainable Development Outlook

NATO countries' military expenditures and sustainable development perspectives are essential aspects of international relations and policy analysis. This section focuses on NATO countries' military expenditures and sustainable development outlooks. The ratio of military expenditures to GDP is an important economic indicator that shows how much economic resources countries allocate to the defense sector. This ratio is a key indicator that reflects each country's defense policies, economic priorities, and strategic security approach. The Sustainable Development Index (SDI) shows the performance of countries in economic, social, and environmental areas. The SDI reveals the extent to which countries place sustainability at the center of their development processes and what kind of environment, standard of living, and educational opportunities they leave for future generations. In this scope, data on NATO countries are shown in Figure 1.

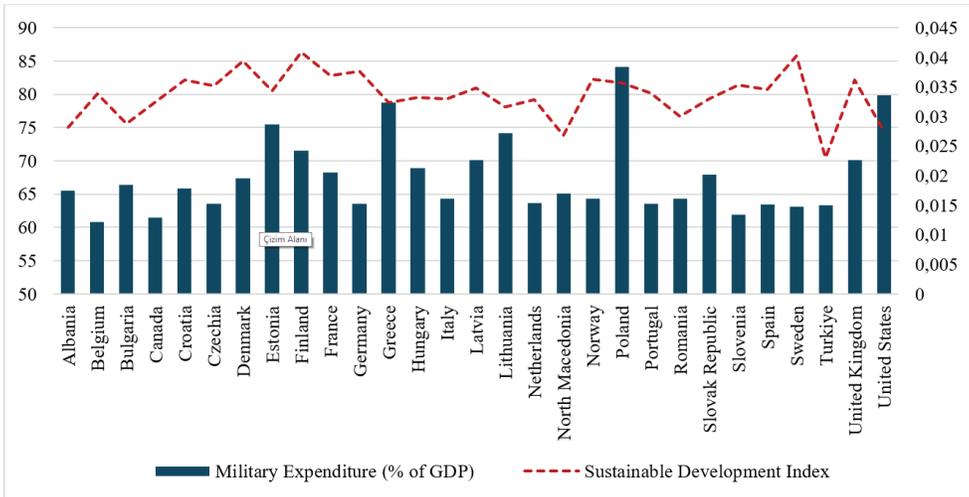

**Figure 1.** Military Expenditures and Sustainable Development Outlook

Figure 1 shows the military expenditures and sustainable development outlook for NATO countries in 2023. The ratio of military expenditures to GDP directly relates to global geopolitical positions, security threat perceptions, and strategic defense policies. Worldwide, these ratios play a key role in determining countries' defense strategies and priorities. For example, Poland has increased its military expenditure to 3.83% of its GDP. This high ratio responds to Poland's growing security needs, especially in Eastern Europe, and its close cooperation with NATO. Similarly, Greece (3.23%) and the United States (3.36%) also have high levels of military expenditures. Greece's high rate can be seen as a response to historical tensions with Türkiye and regional security challenges.[9] As a global superpower, the United States allocates a high proportion of resources due to its numerous international military commitments and military presence worldwide. Meanwhile, countries such as Spain (1.51%), the Netherlands (1.53%), and Belgium (1.21%) are more conservative in their military expenditures. These countries generally have lower threat perceptions and domestic policies emphasizing social rather than military expenditures. The Netherlands and Belgium, in particular, prioritize diplomacy over the military, taking an active role in international peace and cooperation.[10] Türkiye's military expenditures account for 1.50% of its GDP. This ratio shows that Türkiye adopts a strategic approach to both internal security problems and external threats due to its geopolitical position. Türkiye feels the need to keep its military capacity strong, primarily due to its proximity to hot conflict zones such as the Eastern Mediterranean and Syria.[11]

The SDI shows how central sustainability is to the development process and how successful countries are in providing a healthy environment, high living standards, and equal educational opportunities for future generations. Finland (86.35) and Sweden (85.7) have the highest scores on sustainable development. These scores reflect their success in investing in environmentally friendly technologies, improving energy efficiency, and maintaining high educational standards. Denmark (85) is also among the leading countries in sustainable development with a similarly high score. The common feature of these countries is that they have developed policies supporting the transition to green energy, waste management, and social equity.[12] Moreover, large European economies such as France (82.76) and Germany (83.45) score in the middle range. These countries are taking important steps to improve the quality of urban planning and social services while trying to balance industrialization and economic growth with environmental sustainability. Norway (82.23) and the Netherlands (79.21) also fall into this category and emphasize sustainable development policies, especially by keeping environmental standards high. In contrast, Türkiye (70.47) and North Macedonia (73.8) score lower in the index. This indicates that these countries have not made sufficient progress in areas such as environmental management, education, and health care. The United States (74.43) is another country facing severe challenges to sustainable development and scoring lower than expected, mainly due to its environmental policies and social inequalities.[13]

Providing data on military expenditures and the SDI, Figure 1 shows that some countries score high on the SDI despite their high military expenditures. This points out that countries with large economies are able to balance their high military expenditures with investments in other areas. In contrast, countries like Sweden and Finland score very high on the SDI (85.7 and 86.35, respectively) while limiting their military expenditure (1.47% and 2.42% of their GDP, respectively). These countries optimize their development by directing their resources to areas such as education, health, and environmental sustainability instead of defense. Countries like Türkiye, while maintaining military expenditure at 1.50% of their GDP, score lower on the SDI with 70.47. This perhaps reflects the diversion of economic resources to defense spending as well as the underinvestment in other critical areas, such as environmental management and social services. Some smaller economies, such as North Macedonia, have both low military expenditures (1.70% of its GDP) and low SDI (73.8). This suggests that overall economic and resource constraints make it difficult for the countries to make progress in both areas.

---

9 Dionysios Chourchoulis, "Greece, Cyprus and Albania", *The Handbook of European Defence Policies and Armed Forces*, 2018, pp. 313-329; Nimantha Manamperi, "Does Military Expenditure Hinder Economic Growth? Evidence from Greece and Turkiye", *Journal of Policy Modeling,* 38:6, 2016 pp. 1171-1193.
10 Liu Geng et al., "Do Military Expenditures Impede Economic Growth in 48 Islamic Countries? A Panel Data Analysis with Novel Approaches", *Environment, Development and Sustainability*, 2023, pp. 1-35.
11 SIPRI Military Expenditure Database (Stockholm International Peace Research Institute, 2024), https://www.sipri.org/databases/milex, accessed 28.06.2024.
12 Bartosz Bartniczak and Andrzej Raszkowski, "Implementation of the Sustainable Cities and Communities Sustainable Development Goal (SDG) in the European Union", *Sustainability,* 14:24, 2022, p. 16808.
13 The Sustainable Development Index Database, 2024, https://www.sustainabledevelopmentindex.org/time-series, accessed 25.06.2024.

Consequently, the relationship between military expenditures and sustainable development varies from country to country, depending on regional characteristics, economic structure, political stability, international relations, and societal priorities. This diversity reflects each country's internal dynamics and external factors rather than a specific model or a consistent trend. NATO countries show a wide geographical and economic diversity that differentiates the impacts of military expenditures on sustainable development. Therefore, this study empirically analyzes the impacts of military expenditures on sustainable development in NATO countries. The study aims to develop a perspective on how military expenditures can impact not only national security and defense but also economic development, environmental sustainability, and social welfare. Thus, it reveals how military expenditures directly and indirectly impact sustainable development and tries to explain the differences among NATO countries.

## 2. Literature Review

The limited number of studies on the impact of military expenditures on sustainable development in the existing literature points out the gap in this field. However, there is extensive literature on the relationship between military and defense expenditures and the determinants of development. Therefore, the literature review is organized into four main thematic sections.

The first part of the literature review examines the studies on military expenditures and income inequality. Schwuchow[14] found that military expenditures have a significant impact on income inequality in his study covering 82 countries, both developed and developing countries. Similarly, Ali[15] finds that military expenditures positively impact income inequality by using macroeconomic variables such as economic growth and the size of the armed forces. These studies are consistent with Graham and Mueller's[16] study on Organization for Economic Cooperation and Development (OECD) countries. Graham and Mueller found that military expenditures increased income inequality between 1990 and 2007. Vadlamannati's[17] study of four economies in South Asia shows that military expenditures increase income inequality in times of war, while this effect weakens in times of peace. Using long-run data for Türkiye, Elveren[18] finds causality between military expenditures and income inequality and that these variables are cointegrated. Sharif and Afshan[19] analyze the impact of military expenditures

---

14 Soeren C. Schwuchow, "Military Spending and Inequality in Autocracies: A Simple Model", *Peace Economics, Peace Science and Public Policy,* 24:4, 2018, p. 14.
15 Hamid E. Ali, "Military Expenditures and Inequality: Empirical Evidence from Global Data", *Defence and Peace Economics*, 18: 6, 2007, pp. 519-535.
16 Jeremy C. Graham and Danielle Mueller, "Military Expenditures and Income Inequality among a Panel of OECD Countries in the Post-Cold War Era, 1990-2007", *Peace Economics, Peace Science and Public Policy,* 25:1, 2019, p. 25.
17 Krishna Chaitanya Vadlamannati, "Exploring the Relationship between Military Spending & Income Inequality in South Asia", *William Davidson Institute at the University of Michigan, William Davidson Institute Working Papers Series,* Paper Number 918, 2008, p. 13.
18 Adem Y. Elveren, "Military Spending and Income Inequality: Evidence on Cointegration and Causality for Turkey, 1963-2007", *Defence and Peace Economics* 23: 3, 2012, pp. 289-301.
19 Arshian Sharif and Sahar Afshan, "Does Military Spending Impede Income Inequality? A Comparative Study of on income inequality between India and Pakistan, two rival South Asian powers, and find similar results. In contrast, Zhang et al.[20] find that defense expenditures reduce income inequality in China, but this effect varies across different regions. Michael and Stelios[21] find that military expenditure reduces income inequality in NATO countries between 1977 and 2007. Similarly, Ali[22] finds that military spending reduces income inequality in the Middle East and North Africa.

The second part of the literature review analyzes the studies on the environmental impact of military expenditures. The studies examine the impact of military expenditures on air quality and carbon dioxide emissions for different countries and periods. Noubissi Domguia and Poumie[23] find that defense spending has generally positive effects on air quality in 54 countries from 1980 to 2016. This suggests the potential for military spending to improve certain environmental conditions. In contrast, Kwakwa[24] shows that public and military expenditures significantly increased carbon dioxide emissions in Ghana from 1971 to 2018. Similarly, Erdoğan et al.[25] examine the impact of defense expenditures on carbon emissions in Greece, France, Italy, and Spain and find that these expenditures increase emissions at national and regional levels. Another study by Ahmed et al.[26] found that defense expenditures in OECD countries between 1971 and 2020 had a negative impact on the environment in general and increased carbon dioxide emissions in particular. However, there is also evidence that reducing these expenditures would not adversely impact economic growth. Finally, a large-scale study by Elgin et al.[27] in 160 countries showed a positive relationship between the size of military spending and air pollution. This study provides evidence that high military spending can negatively affect environmental sustainability. Collectively, these studies have produced complex results on the environmental impacts of military expenditures, with positive impacts in some cases but negative impacts dominating in most cases. This is a factor to consider when assessing the environmental dimension of defense policies.

The third part of the literature review examines the studies on the social impacts of military expenditures. These studies reveal various findings on the impact of military expenditures on health, education, and general social welfare. In their study of 90 countries for the period between 1989 and 1998, Aizenman and Glick[28] found that defense expenditures generally have a positive impact on economic growth. They also concluded that defense

---

Pakistan and India", *Global Business Review,* 19:2, 2018, pp. 257-279.
20 Ying Zhang, Rui Wang, and Dongqi Yao, "Does Defence Expenditure Have a Spillover Effect on Income Inequality? A Cross-Regional Analysis in China", *Defence and Peace Economics,* 28:6, 2017, pp. 731-749.
21 Chletsos Michael and Roupakias Stelios, "The Effect of Military Spending on Income Inequality: Evidence from NATO Countries", *Empirical Economics,* 58:3, 2020, pp. 1305-1337.
22 Hamid E. Ali, "Military Expenditures and Inequality in the Middle East and North Africa: A Panel Analysis", *Defence and Peace Economics,* 23:6, 2012, pp. 575-589.
23 Edmond Noubissi Domguia and Boker Poumie, "Economic Growth, Military Spending and Environmental Degradation in Africa", 2019, MPRA Paper No. 97455.
24 Paul Adjei Kwakwa, "The Effect of Industrialization, Militarization, and Government Expenditure on Carbon Dioxide Emissions in Ghana", *Environmental Science and Pollution Research,* 29:56, 2022, p. 85229-85242.
25 Seyfettin Erdogan et al., "Does Military Expenditure Impact Environmental Sustainability in Developed Mediterranean Countries?", *Environmental Science and Pollution Research,* 29:21, 2022. p.p 31612-31630.
26 Zahoor Ahmed et al., "The Trade-off between Energy Consumption, Economic Growth, Militarization, and CO 2 Emissions: Does the Treadmill of Destruction Exist in the Modern World?", *Environmental Science and Pollution Research*, 2022, p. 14.
27 Ceyhun Elgin et al., "Military Spending and Sustainable Development", *Review of Development Economics,* 26: 3, 2022, pp. 1466-1490.
28 Joshua Aizenman and Reuven Glick, "Military Expenditure, Threats, and Growth", NBER Working Paper Series, no. w9618. Cambridge, Mass: National Bureau of Economic Research, 2003.

Does Military Expenditure Impede Sustainable Development? Empirical Evidence from NATO Countriesexpenditures stimulate economic growth under high threat levels. Wilkins[29] showed that defense spending positively impacted economic growth in 85 countries between 1988 and 2002 and that this positive impact persisted despite the decline in defense spending after the end of the Cold War. Lin et al.[30] found a positive relationship between military expenditures and education and health expenditures in OECD countries. The study suggests that military spending may synergistically interact with investments in education and health sectors. Zhang et al.[31] examined the impact of military expenditures on social welfare by comparing BRICS (Brazil, Russia, India, China, South Africa) and G7 (Group of Seven) countries. The findings show that military expenditures play a social welfare-enhancing role in developed countries, but this impact is less obvious in developing countries. Destebaşı,[32] in her study on D-8 countries, found that defense investments positively impacted economic growth along with other social investments, such as health and education. Her study emphasizes that defense spending can support not only security policies but also overall social and economic development. These studies demonstrate that military expenditures can have a restraining impact on social expenditures but can also support economic growth and social welfare. Therefore, assessing the social impact of military spending requires a balanced management of both economic and social policies.

The last part of the literature review analyzes the studies investigating military expenditures' impacts on sustainable development. Many studies have examined the relationship between military and defense expenditures and the determinants of development. There is a vast amount of literature in this area, and various studies have evaluated the effects of military expenditures on development indicators such as economic growth, social welfare, and environmental sustainability. However, studies on the direct impact of military expenditures on sustainable development are quite limited. Among these limited studies, Dudzevičiūtė et al.[33] examines the impact of defense expenditures on sustainable development indicators in the Baltic countries together with economic and strategic factors. Their analysis reveals that the increase in defense expenditures in small states such as Lithuania, Latvia, and Estonia significantly affects many sustainable development indicators, including employment rates, the number of research and development (R&D) personnel, per capita income, and environmental taxes. Moreover, Meiling et al.[34] analyzes the relationship between financial liberalization, health expenditures, and military expenditures in Pakistan and their long and short-term impacts on sustainable development. Their study finds that health expenditures and financial liberalization positively impact sustainable development, while military expenditures have a negative impact. These results, supported by the Toda-Yamamoto causality test, suggest that financial liberalization and health expenditure policies significantly affect progress in this area. Finally, Kamali[35] investigates the role of institutional

29 Nigel Wilkins, "Defence Expenditure and Economic Growth: Evidence from a Panel of 85 Countries", *School of Finance and Economics, University of Technology, Sydney PO Box,* 2004.
30 Eric S. Lin, Hamid E. Ali, and Yu-Lung Lu, "Does Military Spending Crowd Out Social Welfare Expenditures? Evidence from a Panel of OECD Countries", *Defence and Peace Economics,* 26: 1, 2015, pp. 33-48.
31 Ying Zhang, Rui Wang, and Dongqi Yao, "Does Defence Expenditure Have a Spillover Effect on Income Inequality? A Cross-Regional Analysis in China", *Defence and Peace Economics,* 28: 6, 2017, pp. 731-749.
32 Emine Destebaşı, "Savunma, Eğitim ve Sağlık Harcamaları Arasındaki Nedensellik Analizi: D-8 Ülkeleri Örneği", *Enderun,* 1:1, 2017, pp. 28-43.
33 Gitana Dudzevičiūtė et al., "An Assessment of the Relationship between Defence Expenditure and Sustainable Development in the Baltic Countries", *Sustainability,* 13:12, 2021, p. 6916.
34 Li Meiling et al., "The Symmetric and Asymmetric Effect of Defense Expenditures, Financial Liberalization, Health Expenditures on Sustainable Development", *Frontiers in Environmental Science,* 10, 2022, p. 23.
35 Sam Kamali, "Military Expenditure, Institutional Quality and the Sustainable Development Goals: Insight into202 Vol: 20 Issue: 48

quality in achieving sustainable development goals and the impact of military spending on this process. The regression analysis reveals that military expenditures do not directly affect progress towards sustainable development goals, but institutional quality contributes significantly to progress towards these goals. These studies show that the impact of military expenditures on sustainable development is complex and multifaceted. The negative impacts of military expenditures are prominent in some cases, while military expenditures can lead to different consequences in other cases when combined with economic or strategic factors. These findings emphasize the need to carefully assess the potential impact of defense policies and expenditures on sustainable development.

## 3. Data and Methodology

### 3.1. Model Specification and Data

This study analyzes the impact of military expenditures on sustainable development in selected NATO[36] countries by using annual data for the period between 1995 and 2019. The data period and countries are determined based on accessibility. The descriptive statistics of the data set are presented in Table 1.

**Table 1.** Descriptive Statistics

| Variables | Symbol | Description | Mean | S. D. | Min. | Max | Source |
|---|---|---|---|---|---|---|---|
| Sustainable Development Index | SDI | index | -0.291 | 0.175 | -0.788 | -0.085 | HCKL |
| Military Expenditures | MILEX | per capita | 3.602 | 0.841 | 1.903 | 5.996 | SIPRI |
| Foreign Direct Investment | FOREIGN | % of GDP | 0.476 | 0.501 | -2.833 | 1.937 | WB |
| Primary Energy Consumption | ENERGY | per capita | 4.574 | 0.252 | 3.838 | 5.111 | EIA |
| Industrial Production Index | INDUSTRIY | index | 1.981 | 0.106 | 1.623 | 2.216 | IMF |

Note: (1) S.D., Min, and Max denote standard deviation, minimum, and maximum, respectively. (2) HCKL, SIPRI, WB, EIA, and IMF indicate data calculated by the Hickel[37] method, Stockholm International Peace Research Institute database, World Bank-World Development Indicators, Energy Information Administration database, and International Monetary Fund database, respectively.

Given the lack of empirical studies in the literature analyzing the impacts of military expenditures on sustainable development, we include possible control variables to identify the factors impacting sustainable development and its components. We compile these control variables from the literature, including Foreign Direct Investment,[38] Primary Energy

---

the Dynamics of a Large-Scale Attempt at Sustainable Development", *Uppsala University Bachelor Thesis*, 2023, p. 14.
36 These countries are Albania, Belgium, Bulgaria, Canada, Croatia, Czechia, Denmark, Estonia, Finland, France, Germany, Greece, Hungary, Italy, Latvia, Lithuania, the Netherlands, North Macedonia, Norway, Poland, Portugal, Romania, Slovak Republic, Slovenia, Spain, Sweden, Türkiye, the United Kingdom, and the United States.
37 Jason Hickel, "The Sustainable Development Index: Measuring the Ecological Efficiency of Human Development in the Anthropocene", *Ecological Economics,* 167, 2020, p. 31.
38 Abdul Rahim Ridzuan, Nor Asmat Ismail, and Abdul Fatah Che Hamat, "Foreign Direct Investment and Trade Openness: Do They Lead to Sustainable Development in Malaysia?", *Journal of Sustainability Science and Management*, 4, 2018, p. 81-100; Karl P. Sauvant and Howard Mann, "Making FDI More Sustainable: Towards an Consumption,[39] and Industrial Production Index.[40] The research model is expressed in functional form as in Equation 1.

$$SDI_{i,t} = f(MILEX_{i,t}, FOREIGN_{i,t}, ENERGY_{i,t}, INDUSTRY_{i,t}) \qquad (1)$$

After logarithms are taken, the research model can be expressed as in Equation 2:

$$SDI_{i,t} = \beta_0 + \beta_1 MILEX_{i,t} + \beta_2 FOREIGN_{i,t} + \beta_3 ENERGY_{i,t} + \beta_4 INDUSTRY_{i,t} + \varepsilon_{i,t} \qquad (2)$$

In Equation 2, $i = 1, \ldots, 29$ denotes each country and $t = 1995, \ldots, 2019$ denotes time. $\beta_0$ and $\beta_1$ denote the intercept and error terms, while $\beta_1, \beta_2, \beta_3,$ and $\beta_4$ denote long-term elasticities.

### 3.2. Methodology

This study consists of an empirical process with five main stages. First, we test for cross-sectional dependence using the Breusch-Pagan LM, Pesaran scaled LM, Bias-corrected scaled LM, and Pesaran CD tests. Second, we test for slope homogeneity using the Slope homogeneity test proposed by Pesaran and Yamagata.[41] Third, we apply the Cross-Sectional Augmented IPS (CIPS) panel unit root test developed by Pesaran[42] to determine each variable's level of integration. Fourth, we use the Durbin-Hausman panel cointegration test to analyze the long-run relationships. Fifth, we apply the panel Augmented Mean Group (AMG) estimator to estimate the long-run parameters. The details of each of these tests are explained in the following.

***Cross-Sectional Dependence Tests:*** In panel data analysis, these tests are used to identify hidden connections between different observations. They fall into two main categories: Pesaran's CD test and the Breusch-Pagan LM test. The Pesaran test measures the overall links between observations, while the Breusch-Pagan test examines the variances of the error terms. The implementation process consists of four main steps: data preparation, model setup, test application, and evaluation of the results. These tests are important to understand the dependence structure between data and to interpret the analysis results correctly. They also identify potential interactions and dependencies in the data set to improve the reliability

---

Indicative List of FDI Sustainability Characteristics", *The Journal of World Investment & Trade*, 20:6, 2019, pp. 916-952.
39 Asma Esseghir and Leila Haouaoui Khouni, "Economic Growth, Energy Consumption and Sustainable Development: The Case of the Union for the Mediterranean Countries", *Energy,* 71, 2014, p. 218-225; Abdeen Mustafa Omer, "Energy, Environment and Sustainable Development", *Renewable and Sustainable Energy Reviews,* 12:9, 2008, pp. 2265-2300.
40 Ibrahim H. Garbie, "An Analytical Technique to Model and Assess Sustainable Development Index in Manufacturing Enterprises", *International Journal of Production Research,* 52:16, 2014, pp. 4876-4915; Marianna Gilli et al., "Sustainable Development and Industrial Development: Manufacturing Environmental Performance, Technology and Consumption/Production Perspectives", *Journal of Environmental Economics and Policy*, 6:2, 2017, pp. 183-203; Ajay Kumar Singh et al., "Assessment of Global Sustainable Development, Environmental Sustainability, Economic Development and Social Development Index in Selected Economies", *International Journal of Sustainable Development and Planning*, 16:1, 2021, pp. 123-138.
41 M. Hashem Pesaran and Takashi Yamagata, "Testing Slope Homogeneity in Large Panels", *Journal of Econometrics*, 142:1, 2008, pp. 50-93.
42 M. Hashem Pesaran, "A Simple Panel Unit Root Test in the Presence of Cross-section Dependence", *Journal of Applied Econometrics,* 22:2, 2007, pp. 265-312.

of econometric analyses.[43] Therefore, we apply Breusch-Pagan LM, Pesaran scaled LM, Bias-corrected scaled LM, and Pesaran CD tests to detect inter-unit dependencies.

***Slope Homogeneity Tests:*** These tests parameter homogeneity between groups in panel data analysis. They check whether the slope coefficients of independent variables differ across groups. They are usually separated into two main categories: the Pesaran and Yamagata[44] test and the Swamy[45] Random Coefficients model. The Pesaran and Yamagata test measures homogeneity of variance across groups, while the Swamy model generates separate coefficient estimates for each group and tests the significance of these estimates. The implementation process includes the steps of data preparation, model setup, testing, and evaluation of the results. These steps include organizing the data, selecting the appropriate model, and calculating test statistics. The results indicate the presence or absence of homogeneity between slope coefficients. In our study, we use the test developed by Pesaran and Yamagata to examine the homogeneity of slope coefficients.

***Panel Unit Root Test:*** To test the stationarity of the series in panel data sets, we use the CIPS (cross-sectionally augmented IPS) test developed by Westerlund,[46] which takes horizontal cross-section dependence into account and provides more robust results. This second-generation panel unit root test efficiently accounts for cross-sectional dependence and reduces the tendency of first-generation tests to over-reject. Moreover, by determining the level of integration of the series, it provides a basic prerequisite for other tests used in time series analysis to provide robust results. The testing process starts with data preparation, and then regression models are constructed for each unit, and cross-sectional terms are added to these models. The CIPS statistic is calculated by averaging the t-statistics of all units, and the results are compared with the critical values to determine the presence of a unit root. The CIPS test improves the robustness of unit root tests, especially for dependent data.

***Panel Cointegration Test:*** The Durbin-Hausman panel cointegration test developed by Westerlund[47] is applied to test for the existence of long-run relationships. This test offers high efficiency in cointegration analysis by actively considering horizontal cross-section dependence and heterogeneity. The Durbin-Hausman test provides reliable results in panel data analyses, especially when some explanatory variables are I(0), making it more robust than other conventional cointegration tests. This test starts with data preparation and continues modeling the relationships between dependent and independent variables by setting up cointegration equations. In the process of applying the test, we determine whether the modeled coefficients are zero and check the stationarity of the residual series.[48]

***Long-Run Elasticities:*** This study uses the Augmented Mean Group (AMG) estimator to estimate the long-run coefficients. The AMG estimator is designed to be robust to cross-sectional dependence and can effectively address various challenges such as heterogeneity, endogeneity, and serial correlation. It is also notable for its usability in the presence of non-stationary variables and provides more reliable and robust estimates while addressing many challenges encountered in panel data analysis.[49] It is also an estimation method that accounts for heterogeneity and dynamic relationships in panel data models. This estimator is mainly used to analyze long-run relationships in macroeconomic and financial data. The AMG estimator takes into account the factors that jointly affect the entire panel with parameters obtained by averaging the group.[50]

---

43 Tsangyao Chang et al., "Renewable Energy and Growth: Evidence from Heterogeneous Panel of G7 Countries Using Granger Causality", *Renewable and Sustainable Energy Reviews,* 52, 2015, pp. 1405-1412; Wei Lan et al., "High Dimensional Cross-Sectional Dependence Test under Arbitrary Serial Correlation", *Science China Mathematics,* 60, 2017, pp. 345-360.
44 M. Hashem Pesaran and Takashi Yamagata, "Testing Slope Homogeneity in Large Panels", *Journal of Econometrics,* 142:1, 2008, pp. 50-93.
45 Paravastu AVB Swamy, "Efficient Inference in a Random Coefficient Regression Model", *Econometrica: Journal of the Econometric Society*, 38:2, 1970, pp. 311-323.
46 Joakim Westerlund, "Panel Cointegration Tests of the Fisher Effect", *Journal of Applied Econometrics,* 23:2, 2008, pp. 193-233.
47 Joakim Westerlund, "Panel Cointegration Tests of the Fisher Effect", *Journal of Applied Econometrics,* 23:2, 2008, pp. 193-233.
48 Salih Turan Katircioglu et al., "Oil Price Movements and Macroeconomic Performance: Evidence from Twenty-Six OECD Countries", *Renewable and Sustainable Energy Reviews,* 44, 2015, pp. 257-270.

## 4. Results and Discussion

The first stage of the empirical analysis is a cross-sectional dependence test for the variables used in the study. The variables analyzed include Sustainable Development Index (SDI), Military Expenditure (MILEX), Foreign Direct Investment (FOREIGN), Primary Energy Consumption (ENERGY), and Industrial Production Index (INDUSTRIY). The four different statistical tests used are as follows: Breusch-Pagan LM test, Pesaran scaled LM test, bias-corrected scaled LM test, and Pesaran CD test. Each test analyzes whether there is a dependence between the cross-sections in the panel data set. The results of the tests are shown in Table 2.

**Table 2.** Cross-Sectional Dependence Test Results

|  | Breusch-Pagan LM | Pesaran Scaled LM | Bias-Corrected Scaled LM | Pesaran CD |
| --- | --- | --- | --- | --- |
| SDI | 4641.66*** | 148.643*** | 148.04*** | 37.68*** |
| MILEX | 2199.05*** | 62.924*** | 62.320*** | 20.08*** |
| FOREIGN | 986.24*** | 20.362*** | 19.76*** | 19.45*** |
| ENERGY | 3287.94*** | 101.134*** | 100.53*** | 21.62*** |
| INDUSTRIY | 5034.90*** | 162.44*** | 161.84*** | 40.39*** |

Note: *, **, and *** denote statistical significance at the 10%, 5%, and 1% levels, respectively.

The results in Table 2 show that all four tests for all variables are statistically significant at the significance level of 1%. This suggests a strong cross-sectional dependence among all the variables analyzed. In other words, there is a significant degree of dependence among the cross-sections of each variable in the panel data set. This dependence indicates that a change in one variable can affect other variables. It also suggests that events occurring in one NATO country can quickly spread to the others. Therefore, to accurately analyze the interactions among NATO countries and the speed of the spread of events, methods that take cross-sectional dependence into account should be used. As a result, the use of econometric models considering such dependencies is of great importance in obtaining valid and robust results.

---

49 Kangyin Dong et al., "Energy Intensity and Energy Conservation Potential in China: A Regional Comparison Perspective", *Energy,* 155, 2018, p. 782-795; Markus Eberhardt and Stephen Bond, "Cross-Section Dependence in Nonstationary Panel Models: A Novel Estimator", 2009, MPRA Paper No. 17692.

50 Dagmawe Tenaw and Alemu L. Hawitibo, "Carbon Decoupling and Economic Growth in Africa: Evidence from Production and Consumption-Based Carbon Emissions", *Resources, Environment and Sustainability*, 6, 2021, pp. 28-51.

**Table 3.** Slope Homogeneity Test Results

| Model | Tests | LM statistics | P-value |
|---|---|---|---|
| $SDI_{i,t}$ | $\widehat{\Delta}$ | 26.424 | 0.000 |
| | $\widehat{\Delta}_{adj}$ | 30.311 | 0.000 |

The second stage of the analysis applies the slope homogeneity test to test the compatibility of the long-run coefficients. The test results shown in Table 3 lead to the rejection of the null hypothesis that the analyzed slopes are homogeneous. This suggests that the differences between the slopes are significant and should be addressed appropriately in the models. The results suggest that methods taking the heterogeneity in the data into account should be used.

**Table 4.** CIPS Panel Unit Root Test Results

| CIPS | Level | | First difference | |
|---|---|---|---|---|
| Variable | Intercept | Intercept & Trend | Intercept | Intercept & Trend |
| SDI | -1.187 | -1.289 | -3.618*** | -4.056*** |
| MILEX | -1.628 | -2.404 | -4.422*** | -4.652*** |
| FOREIGN | -3.383*** | -3.764*** | -5.838*** | -5.954*** |
| ENERGY | -1.650 | -2.858*** | -4.987*** | -5.139*** |
| INDUSTRIY | -1.927 | -1.661 | -3.216*** | -3.323*** |

Note: *, **, and *** denote statistical significance at the 10%, 5%, and 1% levels, respectively.

The third stage of the analysis utilizes the CIPS panel unit root test to check the stationarity of the series. This test provides more reliable results by accounting for cross-sectional dependence. The results in Table 4 include both constant term and constant term and trend cases for five different variables (SDI, MILEX, FOREIGN, ENERGY, INDUSTRIY). The results are presented both in level and first difference. At level results, the CIPS values for most variables are not statistically significant (except FOREIGN, which is statistically significant in both cases). This implies that most variables have unit roots and are non-stationary. However, with the constant term and trend, the ENERGY variable obtains a statistically significant value (-2.858). When the first difference of the variables is taken, all variables show statistically significant results for both "Constant Term" and "Constant Term and Trend". The results indicate that stationarity is achieved when the first differences of the series are taken, and econometric models can be constructed over these stationary series.

**Table 5.** Durbin-Hausman Panel Cointegration

| Statistic | Model | P-value | Decision |
|---|---|---|---|
| $DH_G$ Statistics | 12.315 | 0.012 | Cointegration |
| $DH_P$ Statistics | 4.332 | 0.008 | Cointegration |

Note: ***, **, and * denote the significance at a 1%, 5%, and 10% level, respectively.

The fourth stage of the analysis employs the Durbin-Hausman cointegration test to investigate the long-run relationship between the series. Table 5 shows the results of the Durbin-Hausman panel cointegration test. The test includes $DH_G$ (Durbin-Hausman Group) and DH$_p$ (Durbin-Hausman Panel) statistics. The test value for DH$_G$ is 12.315, and the p-value is 0.012. This p-value indicates that the test is significant at the significance level of 5%; hence, there is cointegration between the variables. Moreover, the test value for DH$_p$ is 4.332, and the p-value is 0.008. This p-value indicates that the test is significant at the significance level of 1% and that there is cointegration between the variables. The results reveal that there is a long-run relationship between the series.

Table 6. Panel AMG Results for All Countries

| Variables | Model | P-value |
|---|---|---|
| MILEX | -0.115 | 0.026 |
| FOREIGN | 0.017 | 0.015 |
| ENERGY | -0.100 | 0.055 |
| INDUSTRIY | -0.253 | 0.054 |

Note: ***, **, and * denote the significance at a 1%, 5%, and 10% level, respectively.

Since we find cointegration between the series in the fourth stage of the analysis, we estimate the long-run coefficients in the fifth stage. We use the AMG estimator to estimate the coefficients. The results in Table 5 provide an overview of the impacts of economic and political factors on the SDI. In NATO countries, military expenditures and the industrial production index have a negative impact on SDI, while FDI has a positive impact. The impact of primary energy consumption is negative and less pronounced than other negative impacts. The results of the study are in line with Dudzevičiūtė et al.[51] and Meiling et al.[52] These results suggest that various economic policies affect sustainable development differently and that these interactions should be carefully evaluated. However, country-specific characteristics need to be considered to reach more robust conclusions. Therefore, country-specific estimates are presented in the next stage of the analysis.

Table 7. Panel AMG Results

| Country | MILEX | FOREIGN | ENERGY | INDUSTRIY |
|---|---|---|---|---|
| Albania | 0.234*** | -0.080* | 0.372*** | 0.142*** |
| Belgium | 0.286* | 0.024 | -0.353 | -0.334* |
| Bulgaria | -0.061**** | 0.001 | -0.293*** | 0.488*** |
| Canada | -0.207 | -0.001 | 2.087** | -0.399* |
| Croatia | -0.174*** | -0.007 | -0.761*** | 0.593*** |
| Czechia | -0.023 | 0.024 | 0.498** | -0.195*** |
| Denmark | -0.009 | 0.024 | 0.148 | -0.623* |
| Estonia | 0.009*** | 0.015 | -0.598*** | -0.629*** |
| Finland | -0.199** | 0.016 | -0.084 | -1.052*** |
| France | -0.463*** | -0.021** | -0.302*** | 0.303*** |
| Germany | -0.932*** | 0.012 | 0.469 | -0.290 |
| Greece | 0.028 | 0.022 | -2.248*** | 0.084 |
| Hungary | -0.054** | 0.010 | -0.189* | 0.361*** |

---

51 Gitana Dudzevičiūtė et al., "An Assessment of the Relationship between Defence Expenditure and Sustainable Development in the Baltic Countries", *Sustainability*, 13:12, 2021, p. 6916.
52 Li Meiling et al., "The Symmetric and Asymmetric Effect of Defense Expenditures, Financial Liberalization, Health Expenditures on Sustainable Development", *Frontiers in Environmental Science*, 10, 2022, p. 23.

| | | | | |
|---|---|---|---|---|
| Italy | -0.013 | -0.008* | -1.043*** | 0.260 |
| Latvia | 0.154*** | 0.053 | 0.491 | -0.563** |
| Lithuania | -0.055*** | 0.071 | 0.505** | -0.410** |
| Netherlands | -0.307** | 0.032 | 0.641** | -2.396*** |
| North Macedonia | -0.079** | 0.061*** | -0.270 | 0.816*** |
| Norway | -0.735*** | -0.046 | -0.079 | 0.834** |
| Poland | -0.243* | 0.036* | -0.745*** | 0.001 |
| Portugal | 0.184 | 0.017 | 0.708*** | -1.141*** |
| Romania | -0.259*** | 0.120*** | -0.331** | 0.604*** |
| Slovak Republic | 0.035 | -0.017 | 0.208 | -0.455*** |
| Slovenia | -0.147** | 0.015 | -0.103 | -0.316*** |
| Spain | -0.115* | 0.046 | -1.149*** | -0.529*** |
| Sweden | 0.110 | 0.029** | 1.220*** | -1.136*** |
| Türkiye | 0.104 | 0.066*** | -0.096 | 0.262 |
| United Kingdom | -0.559*** | -0.029 | -0.542*** | -0.546*** |
| United States | 0.075*** | 0.017 | -1.052*** | -1.083*** |

Note: *, **, and *** denote statistical significance at the 10%, 5%, and 1% levels, respectively.

Table 7 shows that variables such as military expenditures, foreign direct investment, primary energy consumption, and industrial production index have statistically significant effects on the SDI in most of the 26 NATO countries analyzed. Military expenditures significantly impact SDI in 20 countries (79.9%). In many of these countries, the effects are negative. However, in Albania, Belgium, Estonia, Latvia, Estonia, Latvia, and the United States, military expenditures have positive effects on SDI. This result shows that a 1% increase in military expenditures impacts SDI by -0.932-0.286% in NATO countries. FDI has had significant impacts on SDI in eight countries (30.8%), and these impacts are generally positive. Countries with significant positive effects include North Macedonia, Sweden, Türkiye, and Romania. This suggests that a 1% increase in FDI affects the SDI in NATO countries by -0.080-0.120%. Primary energy consumption significantly impacts the SDI in 19 countries (73.1%). In most countries, this effect is negative. However, energy consumption positively affects SDI in Albania, Canada, Czechia, Lithuania, the Netherlands, Portugal, and Sweden. This implies that a 1% increase in primary energy consumption in NATO countries affects SDI by -2.248-2.087%. The industrial production index shows significant effects on SDI in 24 countries (92.3%), and these impacts are generally negative. However, in Albania, Bulgaria, Croatia, Hungary, North Macedonia, Norway, and Romania, the industrial production index has a positive impact on SDI. This result implies that a 1% increase in industrial production impacts SDI by -2.396-0.834% in NATO countries.

The results show that country-specific economic dynamics create significant differences on sustainable development and emphasize that policymakers should develop strategies by taking these interactions into account. From this perspective, it is essential to formulate policies specific to each country's economic conditions.

## 5. Conclusion

This study analyzes the impact of military expenditures on sustainable development in NATO countries between 1995 and -2019. In the analysis process, the Durbin-Hausman cointegration test is used to test the existence of long-term relationships between variables.

In addition, the AMG estimator is employed to estimate the long-run coefficients. These methodological approaches allow us to analyze the potential effects of military expenditures on sustainable development.

The results of the Durbin-Hausman panel cointegration test show that the variables used in the study have a long-run and significant cointegration relationship with SDI. The coefficients of DHG and DHP confirm that there is a significant cointegration relationship among the variables. The AMG estimator results demonstrate that military expenditures generally have a negative effect on SDI in 26 NATO countries, while a positive effect is found in countries such as Albania, Belgium, Estonia, Latvia, Estonia, Latvia, and the United States. This evidence suggests that military expenditures are not only investments in defense and security but may also indirectly impact economic and social development. The impact of Foreign Direct Investment is generally positive and has significant and positive effects on SDI, especially in countries such as North Macedonia, Sweden, Türkiye, and Romania. This result indicates that foreign investment can support sustainable development through economic growth and technological innovation. Primary energy consumption negatively impacts SDI in most countries, while positive effects were found in countries such as Albania, Canada, Czechia, Lithuania, the Netherlands, Portugal, and Sweden. This suggests that increases in energy consumption are directly related to the diversity of energy sources used and their sensitivity to environmental impacts. The industrial production index negatively affects the SDI in most countries analyzed. However, positive effects are observed in some countries, particularly Albania, Bulgaria, Croatia, Hungary, North Macedonia, Norway, and Romania. This result reveals that the impact of industrial activities on sustainable development is closely related to the compliance of industrial processes with environmental standards and sustainability policies.

The results of this study emphasize the need to understand the economic dynamics impacting sustainable development and develop appropriate strategies accordingly. Designing policies specific to each country's economic and social conditions is critical for achieving national and international development goals. Within this scope, more informed and effective policies need to be designed by considering the differences between countries and their unique economic structures. In this context, some policy recommendations have been developed:

(1) Although military expenditures are generally found to have a negative impact on SDI, positive effects are found in some countries. This suggests that defense spending is not limited to security but can also contribute to economic growth through research and technology development investments. Therefore, redirecting a portion of military budgets to research and development can both stimulate the development of defense technologies and have a broad economic impact by enabling the transfer of these technologies to civilian sectors.

(2) Foreign direct investment has significant positive effects on SDI, particularly in some countries. Creating attractive conditions for investors and simplifying investment processes can extend this positive impact. Moreover, integrating environmental and social criteria to direct investments toward sustainable projects allows investments to deliver both economic and social benefits in the long run.

(3) Regarding energy consumption, while negative impacts were observed in most countries, some countries show that energy consumption positively impacts SDI. This suggests that energy policies need to be redesigned. Investing in renewable energy sources and increasing energy efficiency can maximize the positive impacts of energy consumption on sustainable development.

(4) The impact of the industrial production index is negative in most countries analyzed. However, promoting environmentally friendly technologies and integrating environmental standards into industrial processes can positively change these effects. Minimizing the environmental impacts of industrial activities is essential for sustainable development.

In conclusion, these policy recommendations aim to support the achievement of development goals in NATO countries by addressing the various dimensions of sustainable development in a balanced manner. The applicability of each recommendation should be carefully assessed by taking into account country-specificities and the prevailing political and economic conditions.

This study provides important findings on military expenditures and sustainable development but has some limitations. Future research can address these limitations to obtain more comprehensive results. First, this study focuses on sustainable development in general and does not examine the impact of military expenditures on various dimensions of sustainable development. Future research can detail military expenditure's direct and indirect effects on social, economic, and environmental dimensions. This can provide a clearer perspective on the role of military expenditures on sustainable development. Second, only 29 NATO countries with available data are analyzed in this study. However, including non-NATO countries can allow us to assess international impacts on sustainable development from a broader perspective. In particular, including countries with different social, political, and economic structures can help us understand the impact of military expenditures on various forms of governance and development patterns. Finally, larger data sets and various methods of analysis can be used to check the consistency of the findings of this study. Different statistical techniques and modeling approaches can be adopted to enhance the reliability of the results of the current study. Following these guidelines can contribute to a more comprehensive understanding of the relationship between military expenditures and sustainable development.

***Conflict of Interest Statement:***

*The author declares that there is no conflict of interest.*

**REFERENCES**

**Published Works**

AHMED Zahoor, AHMAD Mahmood, MURSHED Muntasir, VASEER Arif I., and KIRIKKALELI Dervis (2022). "The Trade-off between Energy Consumption, Economic Growth, Militarization, and CO 2 Emissions: Does the Treadmill of Destruction Exist in the Modern World?", *Environmental Science and Pollution Research*, 48:4, 1-14.

AIZENMAN Joshua and REUVEN Glick (2003). "Military Expenditure, Threats, and Growth", NBER Working Paper Series, no. w9618, *National Bureau of Economic Research Cambridge, Mass., USA*.

ALI Hamid E. (2007). "Military Expenditures and Inequality: Empirical Evidence from Global Data", *Defence and Peace Economics*, 18:6, 519-535.

ALI Hamid E. (2012). "Military Expenditures and Inequality in the Middle East and North Africa: A Panel Analysis", *Defence and Peace Economics*, 23:6, 575-89. https://doi.org/10.1080/10242694.2012.663578.

BARTNICZAK Bartosz and ANDRZEJ Raszkowski (2022). "Implementation of the Sustainable Cities and Communities Sustainable Development Goal (SDG) in the European Union", *Sustainability*, 14:4, 16-38.

BEKMEZ Selehattin and DESTEK M. Akif (2015): "Savunma Harcamalarında Dışlama Etkisinin İncelenmesi: Panel Veri Analizi", *Siyaset, Ekonomi ve Yönetim Araştırmaları Dergisi*, 3:2, 91-110.

BILDIRICI Melike (2017). "CO2 Emissions and Militarization in G7 Countries: Panel Cointegration and Trivariate Causality Approaches", *Environment and Development Economics,* 22:6, 771-91.

CASSEN R. H (1987). "Our Common Future: Report of the World Commission on Environment and Development", *International Affairs*, 64:1, 126.

CHANG Tsangyao, GUPTA Rangan, INGLESI-LOTZ Roula, SIMO-KENGNE Beatrice, SMITHERS Devon and TREMBLING Amy (2015). "Renewable Energy and Growth: Evidence from Heterogeneous Panel of G7 Countries Using Granger Causality", *Renewable and Sustainable Energy Reviews*, 52, 1405-1412.

CHASEK Pamela S., WAGNER Lynn M., LEONE Faye, LEBADA Ana-Maria, and RISSE Nathalie (2016). "Getting to 2030: Negotiating the Post-2015 Sustainable Development Agenda", *Review of European, Comparative & International Environmental Law,* 25:1, 5-14.

CHOURCHOULIS Dionysios (2018). "Greece, Cyprus and Albania", *The Handbook of European Defence Policies and Armed Forces*, 313-329.

DESTEBAŞI Emine (2017). "Savunma, Eğitim ve Sağlık Harcamaları Arasındaki Nedensellik Analizi: D-8 Ülkeleri Örneği", *Enderun,* 1:1, 28-43.

DONG Kangyin, SUN Renjin, HOCHMAN Gal, and LI Hui (2018). "Energy Intensity and Energy Conservation Potential in China: A Regional Comparison Perspective", *Energy,* 155, 782-795.

DUDZEVIČIŪTĖ Gitana, BEKESIENE Svajone, MEIDUTE-KAVALIAUSKIENE Ieva, and ŠEVČENKO-KOZLOVSKA Galina (2021). "An Assessment of the Relationship between Defence Expenditure and Sustainable Development in the Baltic Countries", *Sustainability,* 13:12, 6916.

EBERHARDT Markus and BOND Stephen (2009). "Cross-Section Dependence in Nonstationary Panel Models: A Novel Estimator", MPRA Paper No. 17692, https://mpra.ub.uni-muenchen.de/id/eprint/17692, accessed 22.05.2024.

ELGIN Ceyhun, ELVEREN Adem Y., ÖZGÜR Gökçer, and DERTLİ Gül (2022). "Military Spending and Sustainable Development", *Review of Development Economics,* 26:3, 1466-1490.

ELVEREN Adem Y. (2012). "Military Spending and Income Inequality: Evidence on Cointegration and Causality for Turkey, 1963–2007", *Defence and Peace Economics,* 23:3, 289-301.

ERDOGAN Seyfettin, GEDIKLI Ayfer, ÇEVIK Emrah İsmail, and ÖNCÜ Mehmet Akif (2022). "Does Military Expenditure Impact Environmental Sustainability in Developed Mediterranean Countries?", *Environmental Science and Pollution Research,* 29:21, 31612-31630.

ESSEGHIR Asma and HAOUAOUI KHOUNI Leila (2014). "Economic Growth, Energy Consumption and Sustainable Development: The Case of the Union for the Mediterranean Countries", *Energy*, 71, 218-225.

GARBIE Ibrahim H. (2014). "An Analytical Technique to Model and Assess Sustainable Development Index in Manufacturing Enterprises", *International Journal of Production Research,* 52:16, 4876-4915.

GENG Liu, ABBAN Olivier Joseph, HONGXING Yao, OFORI Charles, COBBINAH Joana, AMPONG Sarah Akosua, and AKHTAR Muhammad (2023). "Do Military Expenditures Impede Economic Growth in 48 Islamic Countries? A Panel Data Analysis with Novel Approaches", *Environment, Development and Sustainability,* 1-35.

GILLI Marianna, MARIN Giovanni, MAZZANTI Massimiliano, and NICOLLI Francesco (2017). "Sustainable Development and Industrial Development: Manufacturing Environmental Performance, Technology and Consumption/Production Perspectives", *Journal of Environmental Economics and Policy,* 6:2, 183-203.

GOULD Kenneth A. (2007). "The Ecological Costs of Militarization", *Peace Review,* 19:3, 331-334.

GRAHAM Jeremy C. and MUELLER Danielle (2019). "Military Expenditures and Income Inequality among a Panel of OECD Countries in the Post-Cold War Era, 1990–2007", *Peace Economics, Peace Science and Public Policy*, 25:1, 20180016.

Emre AKUSTA
GUERRERO Omar A. and CASTAÑEDA Gonzalo (2022). "How Does Government Expenditure Impact Sustainable Development? Studying the Multidimensional Link between Budgets and Development Gaps", *Sustainability Science,* 17:3, 987-1007.

HICKEL Jason (2020). "The Sustainable Development Index: Measuring the Ecological Efficiency of Human Development in the Anthropocene", *Ecological Economics,* 167, 106331.

HOOKS Gregory and SMITH Chad L. (2004). "The Treadmill of Destruction: National Sacrifice Areas and Native Americans", *American Sociological Review,* 69:4, 558–75.

ISIKSAL Aliya Zhakanova (2021). "Testing the Effect of Sustainable Energy and Military Expenses on Environmental Degradation: Evidence from the States with the Highest Military Expenses", *Environmental Science and Pollution Research,* 28:16, 20487-20498.

JORGENSON Andrew K., CLARK Brett, and KENTOR Jeffrey (2010). "Militarization and the Environment: A Panel Study of Carbon Dioxide Emissions and the Ecological Footprints of Nations, 1970–2000", *Global Environmental Politics,* 10:1, 7-29.

KATIRCIOGLU Salih Turan, SERTOGLU Kamil, CANDEMIR Mehmet, and MERCAN Mehmet (2015). "Oil Price Movements and Macroeconomic Performance: Evidence from Twenty-Six OECD Countries", *Renewable and Sustainable Energy Reviews,* 44, 257-270.

KRAKWA Paul Adjei (2022). "The Effect of Industrialization, Militarization, and Government Expenditure on Carbon Dioxide Emissions in Ghana", *Environmental Science and Pollution Research,* 29:56, 85229-85242.

LAN Wei, PAN Rui, LUO RongHua, and CHENG YongWei (2017). "High Dimensional Cross-Sectional Dependence Test under Arbitrary Serial Correlation", *Science China Mathematics,* 60, 345-360.

LIN Eric S., ALI Hamid E., and LU Yu-Lung (2015). "Does Military Spending Crowd Out Social Welfare Expenditures? Evidence from a Panel of OECD Countries", *Defence and Peace Economics,* 26:1, 33-48.

LOMAZZI Marta, BORISCH Bettina, and LAASER Ulrich (2014). "The Millennium Development Goals: Experiences, Achievements and What's Next", *Global Health Action,* 7:1, 23695.

MANAMPERI Nimantha (2016). "Does Military Expenditure Hinder Economic Growth? Evidence from Greece and Turkey", *Journal of Policy Modeling,* 38:6, 1171-1193.

MEILING Li, TASPINAR Nigar, YAHYA Farzan, HUSSAIN Muhammad, and WAQAS Muhammad (2022). "The Symmetric and Asymmetric Effect of Defense Expenditures, Financial Liberalization, Health Expenditures on Sustainable Development", *Frontiers in Environmental Science,* 10.

MICHAEL Chletsos and STELIOS Roupakias (2020). "The Effect of Military Spending on Income Inequality: Evidence from NATO Countries", *Empirical Economics,* 58:3, 1305-1337.

OMER Abdeen Mustafa (2008). "Energy, Environment and Sustainable Development", *Renewable and Sustainable Energy Reviews,* 12:9, 2265-2300.

PELLOW David Naguib (2007). *Resisting Global Toxics: Transnational Movements for Environmental Justice*. MIT Press.

PESARAN M. Hashem (2007). "A Simple Panel Unit Root Test in the Presence of Cross-section Dependence", *Journal of Applied Econometrics,* 22:2, 265-312.

PESARAN M. Hashem and YAMAGATA Takashi (2008). "Testing Slope Homogeneity in Large Panels", *Journal of Econometrics,* 142:1, 50-93.

RIDZUAN Abdul Rahim, ISMAIL Nor Asmat, and HAMAT Abdul Fatah Che (2018). "Foreign Direct Investment and Trade Openness: Do They Lead to Sustainable Development in Malaysia?", *Editorial Board*, 81, 0-1.

SAUVANT Karl P. and MANN Howard (2019). "Making FDI More Sustainable: Towards an Indicative List of FDI Sustainability Characteristics", *The Journal of World Investment & Trade,* 20:6, 916-952.

SCHWUCHOW Soeren C. (2018). "Military Spending and Inequality in Autocracies: A Simple Model", *Peace Economics, Peace Science and Public Policy,* 24:4.

SHARIF Arshian and AFSHAN Sahar (2018). "Does Military Spending Impede Income Inequality? A Comparative Study of Pakistan and India", *Global Business Review,* 19:2, 257-279.

SINGH Ajay Kumar, JYOTI Bhim, KUMAR Sanjeev, and LENKA Sanjaya Kumar (2021). "Assessment of Global Sustainable Development, Environmental Sustainability, Economic Development and Social Development Index in Selected Economies", *International Journal of Sustainable Development and Planning,* 16:1, 123-138.

SOLARIN Sakiru Adebola, AL-MULALI Usama, and OZTURK Ilhan (2018). "Determinants of Pollution and the Role of the Military Sector: Evidence from a Maximum Likelihood Approach with Two Structural Breaks in the USA", *Environmental Science and Pollution Research,* 25:31, 30949-30961.

Does Military Expenditure Impede Sustainable Development? Empirical Evidence from NATO Countries

SWAMY Paravastu AVB (1970). "Efficient Inference in a Random Coefficient Regression Model", *Econometrica: Journal of the Econometric Society*, 311-323.

TENAW Dagmawe and HAWITIBO Alemu L. (2021). "Carbon Decoupling and Economic Growth in Africa: Evidence from Production and Consumption-Based Carbon Emissions", *Resources, Environment and Sustainability,* 6, 100040.

VADLAMANNATI Krishna Chaitanya (2008). "Exploring the Relationship between Military Spending & Income Inequality in South Asia", *William Davidson Institute Working Paper*, 918, 56-75.

WESTERLUND Joakim (2008). "Panel Cointegration Tests of the Fisher Effect", *Journal of Applied Econometrics,* 23:2, 193-233.

WILKINS Nigel (2004). "Defence Expenditure and Economic Growth: Evidence from a Panel of 85 Countries", *School of Finance and Economics, University of Technology, Sydney PO Box 123*.

ZHANG Ying, WANG Rui, and YAO Dongqi (2017). "Does Defence Expenditure Have a Spillover Effect on Income Inequality? A Cross-Regional Analysis in China", *Defence and Peace Economics,* 28:6, 731-749.

**Internet Sources**

KAMALI Sam (2023). "Military Expenditure, Institutional Quality and the Sustainable Development Goals: Insight into the Dynamics of a Large-Scale Attempt at Sustainable Development", Uppsala University Bachelor Thesis, 2023, p. 14. https://www.diva-portal.org/smash/record.jsf?pid=diva2:1742856, accessed 26.05.2024.

NOUBISSI DOMGUIA Edmond and POUMIE Boker (2019). "Economic Growth, Military Spending and Environmental Degradation in Africa", MPRA Paper No. 97455. https://mpra.ub.uni-muenchen.de/id/eprint/97455, accessed 15.05.2024.

"SIPRI Military Expenditure Database", Stockholm International Peace Research Institute, 2024. https://www.sipri.org/databases/milex, accessed 01.05.2024.

"The Sustainable Development Index Database", 2024, https://www.sustainabledevelopmentindex.org/time-series, accessed 14.05.2024.